# Strong-ARM Dynamic Latch Comparators: Design and Analyses on CAD Platform

Kasi Bandla[1], Dipankar Pal[2]

[1&2]Department of Electrical and Electronics Engineering, BITS-Pilani, K.K Birla Goa Campus, Goa, India.
[1]p20170429@goa.bits-pilani.ac.in

***Abstract -*** Strong-ARM Dynamic Latch Comparators are widely used in high-speed analog-to-digital converters (ADCs), sense amplifiers in memory, RFID applications, and data receivers. This paper presents different methods to improve the performance of Strong-Arm latch-based comparators. The comparator's significant features such as power dissipation, propagation delay, offset voltage, clock feedthrough, area, and kickback noises are discussed and compared with state-of-the-art candidate topologies. Simulation results show that the new comparator topologies of Strong-ARM Dynamic Latch proposed by these authors gave the best results. The proposed designs are tested. The simulations are carried out using UMC 180nm double metal, double poly standard CMOS process technology, for a 100 MHz clock, at 1.8V supply-rail on the Cadence Virtuoso EDA platform.

## 1. Introduction

In recent years, the integrated circuit (IC) design industry has reported significant interest in analog-to-digital converters (ADC). Demand for ancillary low-power, high-speed building blocks, and design methodologies increases with the development of portable electronic systems, wireless communication devices, consumer electronics, and medical equipment. This results in the integration of conventional ADCs with several functional blocks within a single wafer area to produce high-speed systems with low power consumption. Yet a few ADC features like smaller transistor sizes, low power dissipation, and high-speed operation, are difficult to meet simultaneously [1]. A new approach in the design of ADC is therefore required at low supply rails with optimum transistor dimensions [2].

The crucial element in the ADC design process that controls the accuracy and speed of converters is the comparator. The need for high-speed, high-resolution, and low-power comparators exists for switching power regulators, data receivers, memory circuits, radio frequency identification (RFID), and other devices [3]. In order to amplify low input swing quickly and regeneratively to a large value, high-performance comparators are required. A high gain and large bandwidth are therefore necessary for a fast comparator to operate with resolution and accuracy [4]. CMOS dynamic latch comparators are widely used in a variety of applications due to their high input impedance, full output swing, and high speed. By using a positive feedback mechanism in the regeneration mode, these dynamic latches can report improvements. However, such a latched comparator for low-voltage operations can minimize dynamic input ranges and a comparable differential mechanism can occasionally increase the power dissipation [5-6]. The random offset voltages produced by latch-type comparators also might reduce their precision due to device mismatches and random noise. As a result, reducing noise and offset voltages is one of the critical design challenges for the dynamic latched comparator design that restricts speed [7]. To reduce the offset voltage, a pre-amplifier is typically used before the regenerative latch stage. This pre-amplifier can amplify a small input signal to a large output signal, helping it overcome kickback noise and latch offset voltage [8]. Nonetheless, excessive static power dissipation due to additional circuit components makes a pre-amplifier-based comparator unattractive. Charge Sharing Dynamic Latch Comparator (CSDLC) addressed the static power consumption problem [9]. However, it is unable to offer rail-to-rail output swing during either clock cycle. Further, as both output nodes are transitioning at positive and negative CLK edges, the circuit's average dynamic power consumption is also higher. Strong-ARM Latch-based comparator architecture is one of the most widely used Dynamic Latch Comparator (DLC) architectures. Its zero static power dissipation, high input impedance, rail-to-rail outputs, and comparatively low input-referred offset voltage are some of the features contributing to its acceptance as a design of choice [10]. The modified Strong-ARM Dynamic Latch Comparators have been designed to have low power consumption, high speed, low offset, and area efficiency [11-12]. However, they suffer from a high power delay product and are high on-chip real estate estimates.

Against these backdrops, this work presents some novel comparator architectures based on Strong-ARM latch. When compared to the traditional dynamic latch comparators, the proposed designs are capable of producing high-speed, high resolution with low power consumption at low supply voltages. Section 2 of this paper discusses the fundamentals of operation while Section 3 presents the design approach and design considerations. Simulation results are included in Section 4 along with comparisons with other candidate designs for benchmarking. The paper is concluded in Section 5.

## 2. Principle of Operation

A comparator compares two instantaneous analogue voltages to reflect the polarity of the input difference and generates a "1" or a "0" as the result of the comparison. The general symbol of the comparator can be seen in Fig. 1.

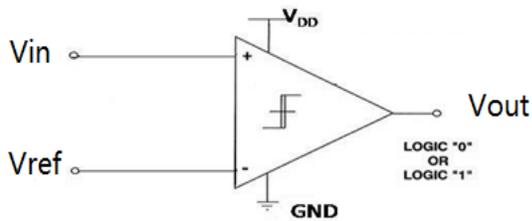

Fig.1 Comparator Symbol

The 'Strong-ARM' comparator is a pair of regenerative latches at the output layered on top of a dynamic differential input gain level. It achieves quick decision-making due to strong positive feedback made possible by two cross-coupled inverter latch pairs and reports low input offset made possible by the matched input differential pair stage [10], [13-15].

In this study, novel Strong-ARM comparator topologies are shown to outperform traditional Strong-ARM architecture in speed, offset, power, and reduced clock feed-through.

## 3. Proposed Design and Methodology

Centered on the Modified Strong-ARM Dynamic Latch Comparator (MSADLC), the following new comparator topologies are proposed here.

### 3.1. Design-1: MASADLC with cascode transistor

Fig. 2 depicts the Strong-Arm Dynamic Latch-based Comparator. It alters the earlier MSADLC topology [11] by incorporating cascode transistors M12 and M13 on both arms above the input transistors to increase the gain. The initial voltage regenerated by the inverter latch is high and also the delay is greatly decreased, which results in the optimization of the power delay product (PDP).

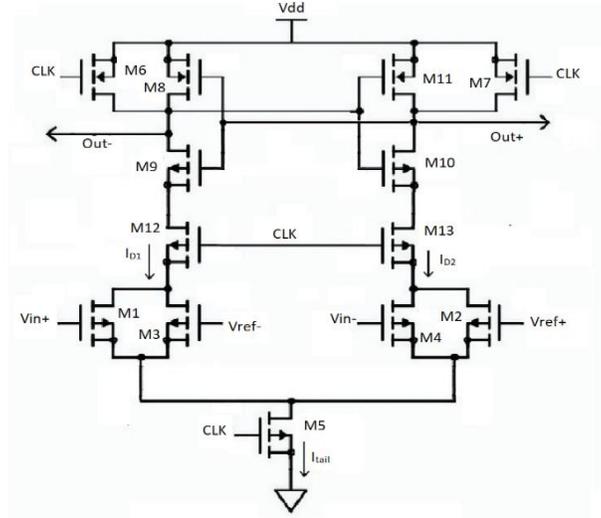

Fig.2 MSADLC with cascode transistor

### 3.1.1 The Fig. 2 operates in two phases.

- **Reset Phase:** When $clk$ signal reaches zero, M5 is turned OFF and the current path is cut off. M6 and M7 reset the differential output nodes, $Vout+$ and $Vout-$ to $VDD$.

- **Comparison Phase:** When $clk$ is high, the cascode transistors M12 and M13 turn ON and the differential output nodes ($Vout+$ and $Vout-$) are isolated from $VDD$. Depending upon the difference between Vin and Vref, cross-coupled inverter pairs made of M8, M9, and M10, M11 regeneratively amplify the difference and determine which of the outputs goes to $VDD$ and which to GND.

### 3.1.2 Design Parameters

*a. Transistor sizes*

In the design of proposed topologies for simulation (as well as for comparisons with other candidate designs for benchmarking) the smallest possible transistor sizes are used to meet the requirements for high speed and low parasitic capacitance. Therefore, digital scaling methodologies are applied to the proposed designs. Table I lists the optimized transistor sizes for each topology from the UMC-180nm model library.

**Table 1. Transistor aspect ratio**

| Transistor Aspect Ratios for Proposed Comparators Design | | | | | |
|---|---|---|---|---|---|
| **Design-1: MSADLC with Cascode** | | **Design-2: MSADLC with Pseudo-NMOS** | | **Design-3: MSADLC with Cascode** | |
| *Transistor* | *W/L* | *Transistor* | *W/L* | *Transistor* | *W/L* |
| M1-M5, M9-M10, M12 and M13 | 720nm /180nm | M1-M5 | 720nm/180nm | M1-M5 and M13-M14 | 720nm/180nm |
| M6-M8 and M11 | 1.13um/180nm | M6-M8 and M11 | 1.13um/180nm | M6-M8 and M11 | 1.13um/180nm |
| --- | --- | M9 and M10 | 240nm/180nm | M10 and M11 | 360nm/180nm |

   *b. Delay analysis*

Two components make up the delay of the comparator: $t_0$ and $t_{latch}$. The first term $t_0$ represents a time to discharge the load capacitance $C_L$ before the first pMOS transistor switches ON. If $clk = 1$, the tail transistor M5, cascode transistors M12 and M13 are ON, and if $|V_{in}^+| > |V_{in}^-|$, transistors $M_1$, $M_3$ accelerate the discharge of $V_{out}^-$ (Fig. 2), turning ON transistor $M_{12}$. This allows us to determine the delay $t_0$ by:

$$t_0 = \frac{C_L \cdot V_{thp}}{I_{D1}} \cong 2 \cdot \frac{C_L V_{thp}}{I_{tail}} \quad (1)$$

where the comparator branch currents are $I_{D1}$ and $I_{D2}$, which together make the total current $I_{tail}$ (= $I_{D1}$ + $I_{D2}$), the threshold voltage of PMOS transistor is $V_{thp}$ and load capacitance is $C_L$.

The drain current $I_{D1}$ in (1) can be approximated to be constant and would be equal to half of the tail current for low input differential voltage ($\Delta V_{in}$). The overall latching delay of two cross-coupled inverters is represented by the second term, $t_{latch}$. From the initial voltage difference $\Delta V_o$, it is anticipated that the end output will be half of the supply rail ($\Rightarrow \Delta V_{out} = V_{DD}/2$).

The latch comes after the comparator, which raises the differential output voltage to its maximum rail-to-rail level. Equation (2) gives the calculation needed to determine the latch assessment time ($t_{latch}$) [13-15]. The delay, $t_{latch}$, is a logarithmic function that depends on the initial output voltage difference at the start of the regeneration phase. (i.e., at t = t$_0$).

$$t_{latch} = \frac{C_L}{g_{m(eff)}} \cdot \ln\left(\frac{\Delta V_{out}}{\Delta V_o}\right)$$

$$t_{latch} \cong \frac{C_L}{g_{m(eff)}} \cdot \ln\left(\frac{V_{DD}/2}{\Delta V_o}\right) \quad (2)$$

In (2), $g_{m(eff)}$ represents the effective transconductance of cross-coupled inverters. The determination of initial differential voltage $\Delta V_o$ using (1) is:

$$\Delta V_o = |V_{out(t=to)}^+ - V_{out(t=t0)}^-|$$

$$\Rightarrow \Delta V_o = |V_{thp}| - \frac{I_{D2} \cdot t_o}{C_L} \quad (3)$$

The differential input current ($\Delta I_{in}$) between the two branches is significantly less than the actual currents (i.e., $I_{D1}$ and $I_{D2}$), which can be approximated by $I_{tail}/2$. As a result, (3) can be rewritten as:

$$\Delta V_o = |V_{thp}|\left(\frac{\Delta I_{in}}{I_{D1}}\right) \cong 2 \cdot |V_{thp}|\left(\frac{\Delta I_{in}}{I_{tail}}\right)$$

solving which, we can get:

$$\Delta V_o = 2 \cdot |V_{thp}| \sqrt{\frac{\beta_{1,2}}{I_{tail}}} \cdot \Delta V_{in} \quad (4)$$

where $\beta_{1,2}$ stands for the input current factor of the transistors and is given by

$$\beta_{1,2} = \mu_n C_{ox} \left(\frac{W}{L}\right)_{1,2} \text{ in } \frac{(\mu)A}{V^2}$$

The supply voltage and input common-mode voltage both affect the tail current, $I_{tail}$. The total delay can be calculated by substituting $\Delta V_o$ from (4) into (2) and by substituting the value of $t_0$ from (1). The results are illustrated in (5) to calculate the total delay.

$$t_{total} = t_0 + t_{latch}$$

$$\Rightarrow t_{total} = 2 \cdot \frac{C_L V_{thp}}{I_{tail}} + \frac{C_L}{g_{m(eff)}} \cdot \ln\left(\frac{V_{DD}/2}{\Delta V_o}\right) \quad (5)$$

which can be rearranged and the overall analytical latency of the proposed dynamic latch comparator is given in equation (6).

$$t_{total} = 2 \cdot \frac{C_L V_{thp}}{I_{tail}} + \frac{C_L}{g_{m(eff)}} \cdot \ln\left(\frac{V_{DD}/2}{2 \cdot |V_{thp}| \sqrt{\frac{\beta_{1,2}}{I_{tail}}} \cdot \Delta V_{in}}\right) \quad (6)$$

   *c. Power analysis*

Typically, the average power dissipated by the supply voltage over a single comparison time is determined by,

$$P_{Avg} = \frac{1}{T} \int_0^T V_{DD} I_D dt = f_{clk} V_{DD} \int_0^T I_D dt \quad (7)$$

where ID represents the current drawn from the supply voltage (VDD), and fclk is the frequency of the comparator clock.

*d. Offset analysis*

Offset happens due to various mismatched parameters stated below. It can be expressed as the error range at input below which the comparator is unable to detect the specified minimum voltage difference. Due to this, the resolution of the comparator and speed are constrained [16-17]. There are no offset-cancelling techniques introduced in this study. However, due to MOS device mismatches, there is a trade-off between high speed and high accuracy. The effects of offset can be alleviated but not entirely removed. The total offset voltage is determined by the mismatch between the threshold voltage $\Delta V_T$, load resistance $\Delta R_L$, and transistor dimensions $\Delta \beta$, and it is given by (8) for the relevant values of ($V_T$, $R_L$ and $\beta$).

$$V_{os} = \Delta V_T + \frac{V_{gs}-V_T}{2}\left[\frac{\Delta R_L}{R} + \frac{\Delta \beta}{\beta}\right] \qquad (8)$$

Inferred from this equation is the fact that the offset voltage decreases as the common mode voltage decreases.

*e. Kickback noise*

This is the noise that appears at the input due to the output coupling with it [18].

*f. Clock feed through*

Clock feedthrough arises because of the clock signal coupling with the output through device capacitances [19-20].

### 3.2. Design-2 MSADLC with Pseudo NMOS Latch

In Fig.3, the cross-coupled CMOS inverter-based latch is replaced by the Pseudo NMOS latch. This lowered the overall capacitance seen by the input to the latch and minimized the delay gradually by reducing the effort delay [21]. In comparison to their CMOS counterparts, Pseudo-NMOS architectures have non-zero static power dissipation [22]. The latch is disabled when the clock signal is LOW and enabled when the clock signal is HIGH to reduce power losses. Power dissipation only takes place while the clock is HIGH. The NMOS applied to the pre-amplifier stage as a load serves the objective of boosting the gain. The gain is generally represented as $g_m.R_{out}$, where $g_m$ is the trans-conductance of the input transistor(s) and $R_{out}$ is the output impedance. Thus, like Cascode Amplifiers, the extra NMOS device increases gain and will henceforth be referred to as the 'cascode transistor'.

The increased gain ensures that an input difference can be resolved faster by the latch. Rearranging the circuit serves two purposes. Here the feedback is taken from the gate of the cascode transistor. This increases the loop gain of the latch and also results in decreasing time constant (delay). The operation of Fig. 3 is identical to that of Fig.2, hence is avoided here for brevity.

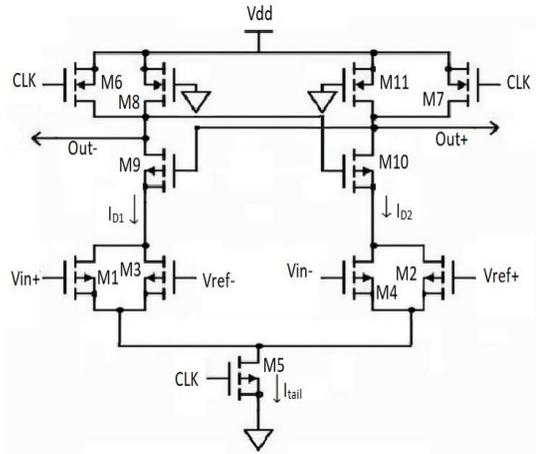

**Fig.3 SADLC with Pseudo NMOS**

### 3.3. Design-3: MSADLC with cascode and pseudo-NMOS

Fig.4 is a further modified architecture, where the feedback is taken from a midpoint of the stack of the differential stage and the latch-stage, above the drain of the cascode-transistor. The input is provided parallel to the transistors driven by $V_{REF}$ latch, and provided above the lower half of the latch constituted by M12-13. Due to the inclusion of cascode transistors above the input differential pair, this topology gives higher gain and the most optimum performance in terms of the design metrics mentioned above. The layout for Fig.4 is shown in Fig.5 on which post layout simulation and Monte Carlo analyses are carried out.

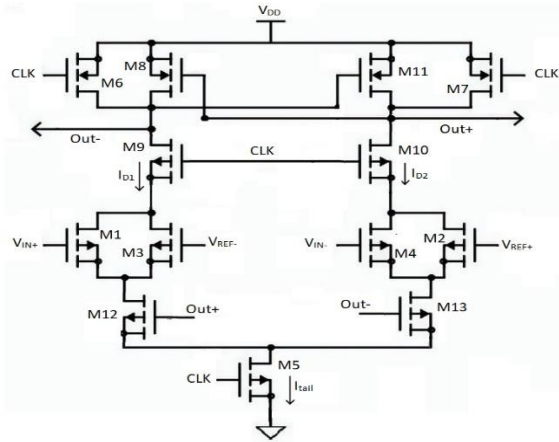

**Fig.4 MSADLC with Cascode and Pseudo-NMOS**

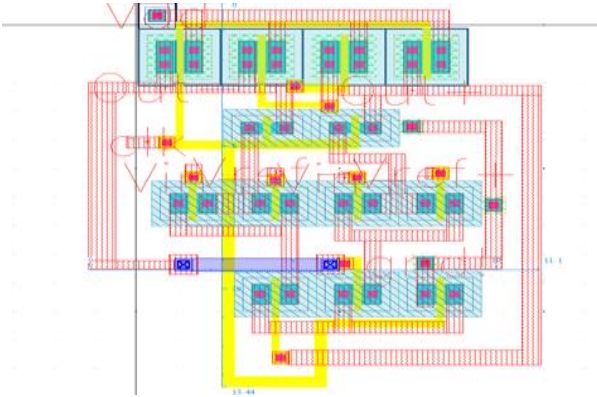

**Fig.5 Layout for design-3 MSADLC with Cascode and Pseudo-NMOS.**

## 4. Results and Discussion

The proposed designs are simulated on Cadence EDA platform using UMC 180nm double metal double poly standard CMOS Technology with $V_{DD}$ = 1.8V. The process entails setting the clock frequency to 100MHz. With newly optimized transistor sizes from Table 1, and on same platform each designed topology is simulated, and significant variables are listed in Table 2. Figures 6 through 9 display typical screen depictions of the simulation results for the proposed Design-3. To avoid redundancy and for the sake of conciseness, the same for the others are omitted.

*4.1. Simulation results*

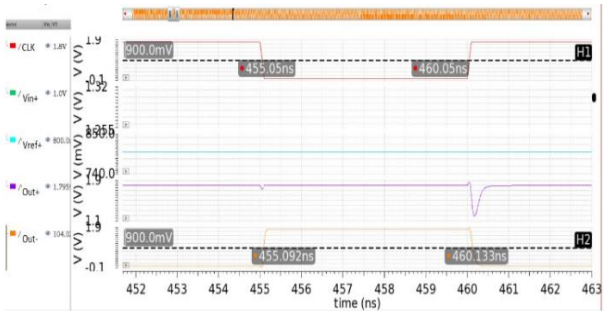

**Fig. 6 Delay measurement**

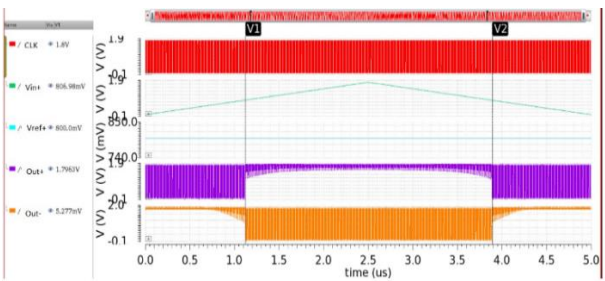

**Fig. 7 Response to a ramp input**

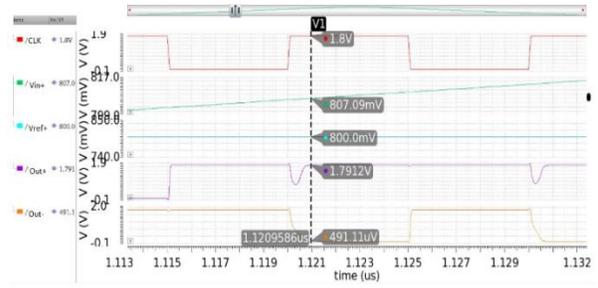

**Fig. 8 Measurement of Offset as input raises**

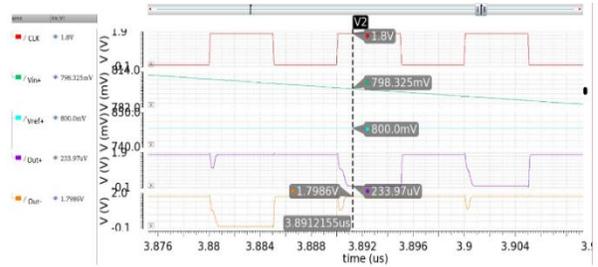

**Fig. 9 Simulation results**

*4.2. Measurement methodology*

The reference voltage $V_{ref+}$ is set at 800 mV, clk = 100MHz, and $V_{DD}$ = 1.8V.

- Power: The input voltage Vin+ is ramped up from 0 V to 1.8 V and down at half the clock period (5 ns). The power dissipation is averaged over the ramp-up/ down period to yield average power.
- Average Delay: The positive input $V_{in+}$ is kept at 1V. The delays tphl and tplh are measured as the clk = 1($V_{DD}$) and Vout- goes to 0; again as the clk =0(0) and $V_{out-}$ goes to $V_{DD}$, respectively. The two delays are averaged over the period to produce the average propagation delay.
- Kickback Noise: The input has been stepped up to 1V after being kept at 600 mV for half the period. The input had a source resistance of 1kΩ, and the excess voltage over this has been measured. The same is done for the negative input, and the two quantities are averaged to get the mean kickback noise.
- Offset Voltage: The input voltage $V_{in+}$ is ramped up/down steadily between 0V to 1.8V. The difference between ($V_{in+}$ - $V_{ref+}$) and the DC–level is the offset voltage seen as the input rises and the output is sampled.
- Clock Feedthrough: When the clk = $V_{DD}$, the input Vin+ is ramped up/ down slowly from 0V to $V_{DD}$. It causes one of the outputs to momentarily shoot up above 1.8V. This deviation is measured as the clock feedthrough. These results are presented in Table 2.

**Table 2. Performance comparisons**

| Topology | Average Delay (ps) | Average Dynamic Power (μW) | PDP (fJ) | Offset Voltage (mV) | Clock-Feed-through (overshoot over 1.8V) (V) | Kickback Noise (overshoot over 1.8V) (V) |
|---|---|---|---|---|---|---|
| *CSDLC** | 178.1 | 18 | 3.2 | 63 | 0.045 | 0.21 |
| *MSADLC** | 93.4 | 4.72 | 0.44 | 6 | 0.097 | 0.005 |
| *Design-1:MSADLC with Cascode* | 62.15 | 4.31 | 0.26 | 2.73 | 0.097 | 0.005 |
| *Design-2:MSADLC with Pseudo NMOS* | 85.6 | 35.07 | 3 | 2.97 | 0.086 | 0.012 |
| *Design-3:MSADLC with Cascode and Pseudo-NMOS* | 62.84 | 4.08 | 0.25 | 2.7 | 0.092 | 0.007 |

*CSDLC: Charge Sharing Dynamic Latch Comparator; *MSADLC: Modified Strong-Arm Dynamic Latch Comparator

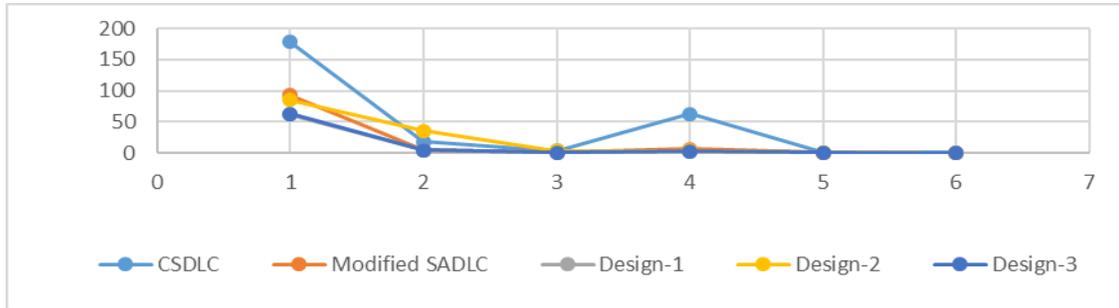

**Fig. 10 Graphical representation of Table 2**

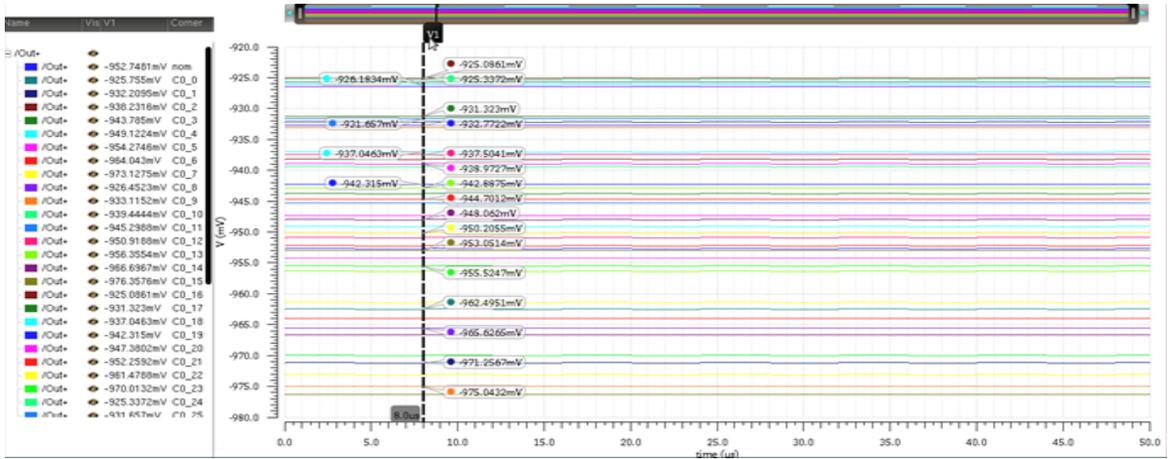

**Fig. 11 Process corner analysis**

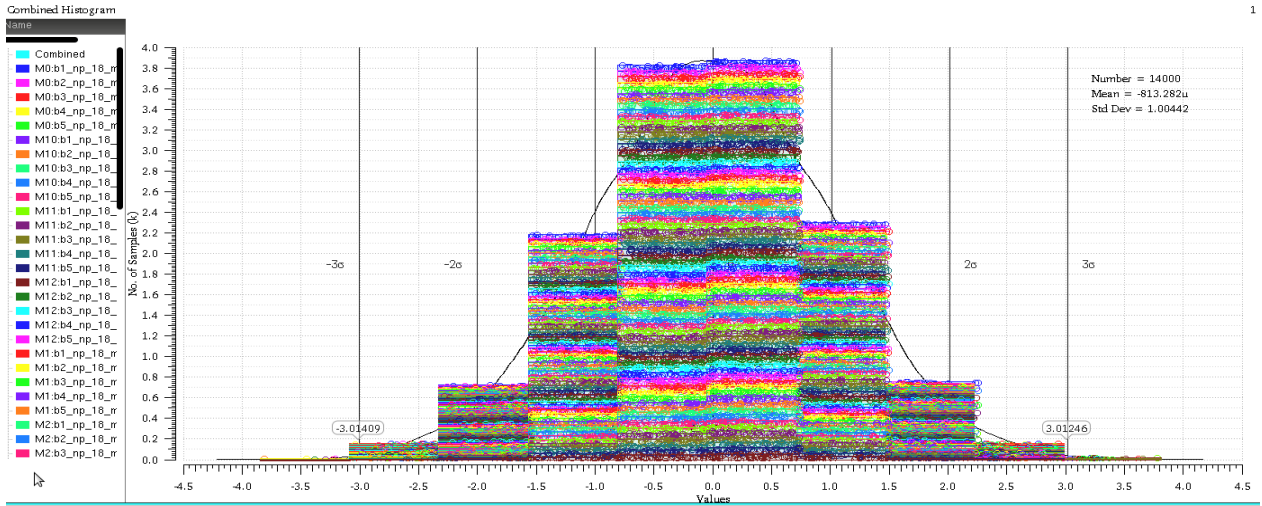

**Fig. 12 Monte-Carlo analysis.**

## 5. Conclusion

In this paper, three new strong-arm dynamic latch-based comparator architectures are proposed. The simulations perfomed on Cadence platform establishes that the proposed comparator architectures are superior in speed, consumes low power and have very low offset as per simulation results done on a supply-rail of 1.8V. Following extraction of the RC from the layout, the simulated outcomes of the comparator circuits were also observed with parasitics, under post layout aspect. The corner analyses and Monte-Carlo analyses have also been performed for each design, although they are not included here for brevity. The estimated area for the design-3 is 11.1μm×13.44 μm as measured from the layout.

Table 2 compares these designs with some candidate designs from open literature for metrics such as delay, kickback noise, average power, clock feed-through, and power delay product (PDP). The Design-2 is performing better when compared to CSDLC. In comparison to MSADLC, design -1 is also better but the design-3 gives best performance of all and shows significant improvement in speed by 32.7 percent; in offset voltage and PDP by 55 percent and 34.2 percent, respectively. As compared to the performance of the MSADLC, the power dissipation and clock feed through is also reduced by 13.5 percent and 5 percent respectively but at the expense of kickback noise which is increased by 40 percent. Without using any offset cancellation techniques, all three designs listed report very low offset. Although design-3 performs best among all, because of their improved performance metrics, these architectures are ideally suited for the design of high-resolution and low-power ADCs.